\documentclass[10pt]{article}

\usepackage{fancyhdr}
\usepackage{extramarks}
\usepackage{amsmath}
\usepackage{amsthm}
\usepackage{amsfonts}
\usepackage{siunitx}
\usepackage{tikz}
\usepackage[plain]{algorithm}
\usepackage{algpseudocode}
\usepackage{multirow}
\usepackage{booktabs}
\usepackage{palatino}
\usepackage{graphicx}
\usepackage{subfigure}
\usepackage[colorlinks,linkcolor=black,anchorcolor=black,citecolor=black,urlcolor=blue]{hyperref}
\usepackage{amsmath,bm}
\usepackage{booktabs}
\usepackage{rotating,longtable}
\usepackage{mathtools}
\usepackage{amssymb}
\usepackage{caption}
\usepackage{capt-of}
\usepackage{mciteplus}
\usepackage{cite}
\usepackage{mathrsfs}
\usepackage[title,titletoc,toc]{appendix}
\usepackage{xr}
\usepackage{parskip}
\usetikzlibrary{automata,positioning}

\renewcommand{\part}[1]{\textbf{\large Part \Alph{partCounter}}\stepcounter{partCounter}\\}

\begin{document}

\title{Mutations on  COVID-19 diagnostic targets
}
\author{ Rui Wang$^1$, Yuta Hozumi$^1$,  Changchuan Yin$^2$ \footnote{Address correspondences to Changchuan Yin. E-mail:cyin1@uic.edu}, and Guo-Wei Wei$^{1,3,4}$ \footnote{Address correspondences to Guo-Wei Wei. E-mail:wei@math.msu.edu} \\
$^1$ Department of Mathematics,
Michigan State University, MI 48824, USA\\
$^2$ Department of Mathematics, Statistics, and Computer Science, \\
University of Illinois at Chicago, Chicago, IL 60607, USA\\
$^3$  Department of Biochemistry and Molecular Biology\\
Michigan State University, MI 48824, USA \\
$^4$ Department of Electrical and Computer Engineering \\
Michigan State University, MI 48824, USA }

\date{}

\maketitle
\begin{abstract}

Effective, sensitive, and reliable diagnostic reagents are of paramount importance for combating the ongoing coronavirus disease 2019 (COVID-19) pandemic at the time there is no preventive vaccine nor specific drug available for severe acute respiratory syndrome coronavirus 2 (SARS-CoV-2). It would be an absolute tragedy if currently used diagnostic reagents are undermined in any manner. Based on the genotyping of 7818 SARS-CoV-2 genome samples collected up to May 1, 2020, we reveal that essentially all of the current COVID-19 diagnostic targets have had mutations. We further show that SARS-CoV-2 has the most devastating mutations on the targets of various   nucleocapsid (N) gene primers and probes, which have been unfortunately used by  countries around the world to diagnose COVID-19. Our findings explain what has seriously gone wrong with a specific diagnostic reagent made in China.  To understand whether SARS-CoV-2 genes have mutated unevenly, we have computed the mutation ratio and mutation $h$-index of all SARS-CoV genes,  indicating that the N gene is the most non-conservative gene in the SARS-CoV-2 genome.  Our findings enable researchers to target the most conservative SARS-CoV-2 genes and proteins for the design and development of COVID-19 diagnostic reagents, preventive vaccines, and therapeutic medicines. 
 
 \end{abstract}

The coronavirus disease 2019 (COVID-19) pandemic outbreak caused by severe acute respiratory syndrome coronavirus 2 (SARS-CoV-2), first reported in Wuhan in December 2019,   has spread to 187 countries and territories with more than 3.481 million infection cases and 244,633 fatalities worldwide by May 1, 2020.  
Additionally, travel restrictions, quarantines, and social distancing measures have essentially put the global economy on hold.   Unfortunately, there is no specific medication nor vaccine for COVID-19 at this moment. Therefore, reopening  economies depends vitally on effective COVID-19 diagnostic testing, patient isolation,  contact tracing, and quarantine. It cannot be overemphasized the importance of diagnostic testing for combating COVID-19.

We reveal that there are many mutations on the COVID-19 diagnostic targets commonly used around the world, including those designated by the United States (US) Centers for Disease Control and Prevention (CDC). These mutations seriously undermine the current global effort in COVID-19 testing, prevention, and control.  Approved by the US Food and Drug Administration (FDA), the CDC has detailed guidelines for COVID-19 diagnostic testing, called  ``CDC 2019-Novel Coronavirus (2019-nCoV) Real-Time RT-PCR Diagnostic Panel'' (\url{https://www.fda.gov/media/134922/download}). The CDC has designated two oligonucleotide primers from regions of the virus nucleocapsid (N) gene, i.e., N1 and N2,  as probes for the specific detection of  SARS-CoV-2. The panel has also selected an additional primer/probe set,  the human RNase P gene (RP), as control samples.  Many other diagnostic primers and probes based on  RNA-dependent RNA polymerase (RdRP, also named  IP2,  IP4, ORF1ab, or  ORF1b), envelope (E), and N genes have been designed  \cite{corman2020detection}  and/or designated by the World Health Organization (WHO) as shown in Table S1 of the Supporting Material,  which provides the details of 41 commonly used diagnostic primers and probes \cite{udugama2020diagnosing}.  
 
 Diagnostic test reagents were designed based on early clinical specimens containing a full spectrum of SARS-CoV-2, particularly the reference genome collected on January 5, 2020, in Wuhan  (SARS-CoV,    NC004718) \cite{wu2020new}. It has been reported that  different primers and probes show nonuniform performance \cite{jung2020comparative,pfefferle2020evaluation,vogels2020analytical,casto2020comparative,nalla2020comparative,chan2020improved}. 

	Our findings are based on the genotyping of 7818 SARS-CoV-2 genome samples collected up to May 1, 2020, which have 5117 single mutations over about 29.8 kilobases (kb).  These mutations occur on all of SARS-CoV-2 genes and proteins, indicating alarming impacts on the current efforts in the development of COVID-19 diagnostic tests, prevention vaccines, and therapeutic medicines. We employ $K$-means methods to cluster these mutations, resulting in globally at least five distinct subtypes of SARS-CoV-2 genomes, from early Cluster I to late Cluster V.  
Table \ref{table:World} shows cluster distributions of samples ($N_{\rm NS}$) and total mutation counts ($N_ {\rm TF}$) for  11 countries. 
 
 \begin{table}[H]
    \centering
    \setlength\tabcolsep{5pt}
    \captionsetup{margin=0.9cm}
    \caption{The cluster distributions of samples ($N_{\rm NS}$) and total mutation counts ($N_ {\rm TF}$) for  11 countries.}
    \begin{tabular}{lcc|cc|cc|cc|cc}
    \hline
      &   \multicolumn{2}{c|}{Cluster I}    & \multicolumn{2}{c|}{Cluster II}    & \multicolumn{2}{c|}{Cluster III}    & \multicolumn{2}{c|}{Cluster IV}   & \multicolumn{2}{c}{Cluster V}     \\ \hline
          Country                &   $N_{\rm NS}$ &$N_ {\rm TF}$  & $N_{\rm NS}$ &$N_ {\rm TF}$  & $N_{\rm NS}$ &$N_ {\rm TF}$  & $N_{\rm NS}$ &$N_ {\rm TF}$  &$N_{\rm NS}$ &$N_ {\rm TF}$ \\  \hline
     US             & 739 &5149  & 248&1623   & 514 & 4968    & 60 &555     & 677&5035  \\
     CA             & 40&240    & 13&72      & 28&193     & 14&119     & 19&126     \\
     AU             & 63&434    & 354&3810   & 182&1315    & 99&873     & 96&691     \\
		UA             & 2&13     & 554&3785   & 607&4206    & 597&5730   & 60&457  \\
		     CN             & 23&54     & 179&865    & 7&63        & 1&13       & 1&7    \\
     DE             & 0 &   0      & 12&42      & 3&18        & 8&70       & 20&131     \\ 
     FR             & 0    & 0     & 14&55      & 105&755     & 6&49       & 66&463    \\
      UK             & 0   & 0      & 23&90      & 10&55       & 4&30       & 0&0        \\
	 IT             & 0       &0   & 6&134      & 22&161      & 12&140     & 0&0        \\
     JP             & 0 &0       & 67&194     & 0&0         & 0&0        & 0&0    \\
     KR             & 0&0       & 26&160     & 0&0         & 0&0       & 0&0    \\ \hline
    \end{tabular}
    \label{table:World}
\end{table}

The US, Canada (CA),  Australia (AU), Ukraine (UA), and China (CN) samples involve all of the five clusters. Among them, China initially had samples only in Clusters I and II and its sample distributions reached to other Clusters after March 2020. Germany (DE) and France (FR) samples are in Cluster II, III, IV, and V. 
 United Kingdom (UK) and Italy (IT) samples are mainly in Clusters II, III, and IV. 
Japan (JP) and Korea (KR) samples belong to Cluster II only.
Cluster II is common to all countries. 

Table  \ref{table:Summary} provides all mutations on various primers and probes and their occurring frequencies in various clusters. More detailed mutation information is given in Tables S2-S42 of the Supporting Material. 
It is interesting to note that N-China-F \cite{udugama2020diagnosing}  is the most inefficient reagent among all primers/probes and its SARS-CoV-2 target has eight mutations involving samples in all five clusters, which may explain many media reports about the inefficiency of certain COVID-19 diagnostic kits made in China. Note that primers and probes  typically have a small length of around 20 nucleotides.

{\small
\begin{table}[H]
\setlength\tabcolsep{2pt}
   \captionsetup{margin=0.5cm}
    \caption{Summary of mutations on COVID-19 diagnostic primers and probes and their occurrence frequencies in clusters. }
 \begin{tabular}{lcccccccc}
    \hline
    Primer/probe  & \# of mutations  & Total frequency & Cluster I  & Cluster II & Cluster III  & Cluster IV  & Cluster V\\ \hline
  	RX7038-N1 primer (Fw)$^a$     & 4   & 5   & 0  & 1   & 3 & 1  & 0     \\
    RX7038-N1 primer (Rv)$^a$     & 6   & 42  & 0  & 28  & 3  & 13 & 0   \\
    RX7038-N2 primer (Fw)$^a$     & 2  & 5   & 0  & 1   & 2  & 1  & 1     \\
    RX7038-N2 primer (Rv)$^a$     & 3  & 9   & 2  & 5   & 2  & 0  & 0     \\
    RX7038-N3 primer (Fw) \cite{nalla2020comparative}  & 5 & 110  & 0  & 98  & 9   & 2  & 1      \\
    RX7038-N3 primer (Rv) \cite{nalla2020comparative}  & 6 & 17   & 1  & 3   & 11  & 1  & 1     \\
    N1-U.S.-P \cite{udugama2020diagnosing}          & 3   & 116    & 1   & 108  & 5   & 2      & 0   \\
    N2-U.S.-P  \cite{udugama2020diagnosing}         & 3   & 31     & 27  & 3    & 2   & 1      & 0   \\
    N3-U.S.-P  \cite{udugama2020diagnosing}         & 8   & 19     & 4   & 5    & 6   & 2      & 2  \\
N-Sarbeco-F$^b$ \cite{corman2020detection}   & 5   & 15   & 3  & 4   & 6  & 0  & 2     \\
	N-Sarbeco-P$^b$\cite{corman2020detection}   & 2   & 3    & 0  & 0   & 1  & 2  & 0     \\
	N-Sarbeco-R$^b$\cite{corman2020detection}   & 7   & 33   & 7  & 6   & 8  & 0  & 12     \\
		 N-China-F \cite{udugama2020diagnosing}          & 8   & 4194   & 9   & 76   & 29  & 4062   & 15  \\
    N-China-R \cite{udugama2020diagnosing}          & 7   & 14     & 2   & 1    & 7   & 3      & 1   \\
    N-China-P  \cite{udugama2020diagnosing}         & 0   & 0      & 0   & 0    & 0   & 0      & 0   \\
		N-HK-F \cite{udugama2020diagnosing}             & 4   & 44     & 0   & 3    & 16  & 25     & 0  \\
    N-HK-R  \cite{udugama2020diagnosing}            & 3   & 12     & 2   & 0    & 7   & 2      & 1  \\
    N-JP-F  \cite{udugama2020diagnosing}            & 2   & 5      & 3   & 2    & 0   & 0      & 0  \\
    N-JP-R    \cite{udugama2020diagnosing}          & 2   & 4      & 0   & 2    & 2   & 0      & 0  \\
    N-TL-F    \cite{udugama2020diagnosing}          & 7   & 40     & 1   & 32   & 4   & 3      & 0  \\
    N-TL-R  \cite{udugama2020diagnosing}            & 6   & 14     & 0   & 6    & 6   & 1      & 1  \\
    N-TL-P  \cite{udugama2020diagnosing}            & 3   & 12     & 0   & 1    & 2   & 9      & 0  \\
	E-Sarbeco-F1$^c$                    & 1   & 1    & 0  & 1   & 0  & 0  & 1     \\   
	E-Sarbeco-R2$^c$                    & 2   & 2    & 1  & 1   & 0  & 0  & 0     \\
  	E-Sarbeco-P1$^c$                    & 2   & 8    & 0  & 6   & 2  & 0  & 0     \\
	E-DE-F   \cite{udugama2020diagnosing}           & 1   & 2      & 0   & 0    & 2   & 0      & 0  \\
  	nCoV-IP2-12669Fw$^c$                & 0   & 0    & 0  & 0   & 0  & 0  & 0     \\
    nCoV-IP2-12759Rv$^c$                & 7   & 39   & 1  & 13  & 22 & 0  & 3     \\
    nCoV-IP2-12696bProbe(+)$^c$         & 1   & 4    & 0  & 0   & 4  & 0  & 0     \\
 	nCoV-IP4-14059Fw$^c$                & 1   & 8    & 0  & 0   & 8  & 0  & 0     \\
 	nCoV-IP4-14146Rv $^c$               & 3   & 13   & 0  & 4   & 5  & 0  & 4     \\
    nCoV-IP4-14084Probe(+)$^c$          & 3   & 9    & 0  & 4   & 5  & 0  & 0     \\
	RdRP-SARSr-F2$^d$                   & 3   & 13   & 0  & 0   & 11 & 0  & 2     \\
	RdRP-SARSr-R1\cite{corman2020detection}$^d$   & 1   & 1    & 0  & 0   & 1  & 0  & 0     \\
	RdRP-SARSr-P2\cite{corman2020detection}$^d$   & 2   & 6    & 0  & 5   & 1  & 0  & 0     \\
     ORF1ab-China-F \cite{udugama2020diagnosing}    & 0   & 0      & 0   & 0    & 0   & 0      & 0   \\
    ORF1ab-China-R \cite{udugama2020diagnosing}     & 0   & 0      & 0   & 0    & 0   & 0      & 0   \\
    ORF1ab-China-P \cite{udugama2020diagnosing}     & 0   & 0      & 0   & 0    & 0   & 0      & 0   \\
    ORF1b-nsp14-HK-F \cite{udugama2020diagnosing}   & 2   & 2      & 0   & 0    & 1   & 0      & 1  \\
    ORF1b-nsp14-HK-R\cite{udugama2020diagnosing}    & 4   & 9      & 0   & 6    & 2   & 2      & 0  \\
    ORF1b-nsp14-HK-P\cite{udugama2020diagnosing}    & 2   & 4      & 1   & 0    & 2   & 1      & 0  \\
	\hline
    \end{tabular}
	$^a$\url{https://www.fda.gov/media/136691/download}\\
    $^b$\url{https://www.eurosurveillance.org/content/table/10.2807/1560-7917.ES.2020.25.3.2000045.t1?fmt=ahah&fullscreen=true}\\
	$^c$\url{https://www.who.int/docs/default-source/coronaviruse/real-time-rt-pcr-assays-for-the-detection-of-sars-cov-2-institut-pasteur-paris.pdf?sfvrsn=3662fcb6_2}\\
	$^d$\url{https://www.who.int/docs/default-source/coronaviruse/protocol-v2-1.pdf?sfvrsn=a9ef618c_2} 
    \label{table:Summary}
\end{table}
}	

Currently, all the primers and probes used in the US target the N gene  \cite{udugama2020diagnosing}. 
Unfortunately, Table  \ref{table:Summary} shows that all of the US CDC designated COVID-19 diagnostic primers have been compromised.  The targets of N gene primers and probes used in Japan, Thailand, and China, including Hong Kong, except for that of  N-China-P, have undergone multiple mutations involving many clusters as well.

It is interesting to note that the targets of four E gene primers and probes have only six mutations. No mutation has been found on the targets of RNA-dependent RNA polymerase-based primers or probes,  nCoV-IP2-12669Fw primer,  ORF1ab-China-F,  ORF1ab-China-R, and  ORF1ab-China-P. However,  the target of  
nCoV-IP2-12759R  recommended by  Institut    Pasteur,   Paris has 7 mutations.  Overall, targets of the envelope and RNA-dependent RNA polymerase based primers and probes have fewer mutations than those of the N gene.  This observation leads us to wonder whether the N gene is particularly prone to mutations.

\begin{table}[H]
    \centering
    \setlength\tabcolsep{5pt}
    \captionsetup{margin=0.9cm}
    \caption{Gene-specific statistics of SARS-CoV-2 single mutations.}
    \begin{tabular}{lccccc}
    \hline
     Gene type & Gene site & Gene length   & Unique SNPs & mutation ratio  &  h-index  \\ 
     \hline
     NSP1      & 266:805       & 540           & 121         & 0.2241          & 8   \\
     NSP2      & 806:2719      & 1914          & 407         & 0.2126          & 16  \\
     NSP3      & 2720:8554     & 5835          & 912         & 0.1563          & 18  \\
     NSP4      & 8555:10054    & 1500          & 203         & 0.1353          & 11  \\
     NSP5(3CL) & 10055:10972   & 918           & 130         & 0.1416          & 10  \\
     NSP6      & 10973:11842   & 870           & 133         & 0.1529          & 8   \\
     NSP7      & 11843:12091   & 249           & 37          & 0.1486          & 5   \\
     NSP8      & 12092:12685   & 594           & 77          & 0.1296          & 4   \\
     NSP9      & 12686:13024   & 339           & 48          & 0.1416          & 6   \\
     NSP10     & 13025:13441   & 417           & 44          & 0.1055          & 4   \\
     NSP11     & 13442:13480   & 39            & 5          & 0.1282           & 2   \\
     RNA-dependent-polymerase  & 13442:16236   & 2796        & 363   & 0.1298  & 15   \\
     Helicase  & 16237:18039   & 1803          & 227         & 0.1259          & 12   \\
     3'-to-5' exonuclease      & 18040:19620   & 1581        & 241   & 0.1524  & 10   \\
     endoRNAse & 19621:20658   & 1038          & 143         & 0.1378          & 10   \\
     2'-O-ribose methyltransferase  & 20659:21552   & 894    & 115   & 0.1286  & 8   \\
     Spike protein     & 21563:25384 & 3819   & 622   & 0.1629  & 17   \\
     ORF3a protein  & 25393:26220   & 825      & 231         & 0.28            & 13   \\
     Envelope protein  & 26245:26472 & 225          & 30          & 0.1333          & 6   \\
     Membrane glycoprotein     & 26523:27191   & 666         & 105   & 0.1577  & 11   \\
     ORF6 protein  & 27202:27387 & 183         & 47          & 0.2568          & 6   \\
     ORF7a protein & 27394:27759 & 363         & 88          & 0.2424          & 6   \\
     ORF7b protein & 27756:27887 & 129          & 10          & 0.0775          & 2   \\
     ORF8 protein  & 27894:28259 & 363          & 90          & 0.2479          & 8   \\
     Nucleocapsid protein  & 28274:29533  & 1257     & 340   & 0.2705  & 29   \\
     ORF10 protein & 29558:29674 & 114       & 27          & 0.2368          & 4   \\
     \hline
    \end{tabular}
    \label{table:MutationStat}
\end{table}

To understand whether there is a differentiation in SARS-CoV-2 gene mutation pattern, we analyze the gene-specific statistics of  SARS-CoV-2 single mutations.  Table \ref{table:MutationStat} lists the mutation ratio, i.e., 
number of unique single-nucleotide polymorphisms (SNPs) over the corresponding gene length, for all SARS-CoV-2 genes. A smaller mutation ratio for a given gene indicates its higher degree of conservativeness. Clearly, ORF7b gene has the smallest  mutation ratio of 0.0775. The N gene has the second largest mutation ratio of 0.2705, which is very close to the largest ratio of 0.2800 for ORF3a gene. To take into the consideration of mutation frequency, we introduce  the mutation $h$-index,  defined as the maximum value of $h$ such that the given  gene section has $h$ single mutations that have each occurred at least $h$ times.  Normally,  larger genes tend to have higher $h$-index.  Table \ref{table:MutationStat} shows that, with a moderate  length, the N gene has the largest $h$-index of 29, which is significantly higher the second largest $h$-index of 18 for NSP3. Therefore, it was truly unfortunate for the world to have selected SARS-CoV-2 N gene primers and probes as diagnostic reagents for combating COVID-19. 

In summary, the targets of currently used COVID-19 diagnostic reagents have had numerous mutations that have seriously undermined our ability to combat  COVID-19.    In the   Supporting Material, we provide a full list of all  5117 SNP variants, including their positions and mutation types.  This information, together with ranking of the degree of the conservativeness of SARS-CoV-2 genes or proteins given in Table \ref{table:MutationStat},  enables researchers to avoid non-conservative genes (or their proteins) and mutated nucleotide segments  in designing COVID-19 diagnosis, vaccine and drugs.

{\bf Methods and materials}
   SARS-CoV-2 genome sequences from infected individuals dated between January 5, 2020, and May 1, 2020, are downloaded from the GISAID database \cite{shu2017gisaid} (\url{https://www.gisaid.org/}). We only consider the records in GISAID with complete genomes and submission dates.  The resulting 7818 complete genome sequences are rearranged according to the reference SARS-CoV-2 genome   \cite{wu2020new} by using the Clustal Omega multiple sequence alignment with default parameters \cite{sievers2014clustal}. 
Gene variants are recorded as single-nucleotide polymorphisms (SNPs). The Jaccard distance   \cite{levandowsky1971distance} is employed to compute the similarities among genome samples. The resulting distance matrix is used in the $k$-means clustering of all samples.

\section{Data Availability} The nucleotide sequences of the SARS-CoV-2 genomes used in this analysis are available, upon free registration, from the GISAID database (\url{https://www.gisaid.org/}). 
 Supporting Material presents  a list of 5117 SNP variants of 7818 SARS-CoV-2 samples across the world, a list of  41 commonly used diagnostic  primers and probes, and tables of mutation details on 41 diagnostic  primers and probes.  The acknowledgments of the SARS-COV-2 genomes are also given in the Supporting Material.  
 
\section*{Acknowledgment}
This work was supported in part by NIH grant  GM126189, NSF Grants DMS-1721024,  DMS-1761320, and IIS1900473,  Michigan Economic Development Corporation,  Bristol-Myers Squibb, and Pfizer. The authors thank The IBM TJ Watson Research Center, The COVID-19 High Performance Computing Consortium, and  NVIDIA for computational assistance.


\end{document}